\documentclass[12pt,a4paper]{article}
\pdfoutput=1

\usepackage{amssymb}
\usepackage{amsmath}
\usepackage{amsfonts}
\usepackage{amsthm}
\usepackage{mathrsfs}
\usepackage{booktabs} 
\usepackage{amsmath} 
\usepackage{tabularx}  
\usepackage{enumitem}
\usepackage{setspace}
\usepackage{color}
\usepackage[pdftex]{graphicx}
\usepackage{authblk}
\usepackage{subfigure}
\usepackage{float}
\usepackage[utf8]{inputenc}
\usepackage{subfigure}  			
\usepackage{adjustbox}
\usepackage[font=small]{caption}	
\usepackage[compress]{cite}         
\usepackage{float}
\usepackage{url}
\usepackage{microtype}
\usepackage{datetime}

\usepackage{pgfplots}   			
	\pgfplotsset{width=10cm}
	\usepackage{bbm}
	\usepackage{tensor}
	\usepackage{slashed}

\numberwithin{equation}{section}

\theoremstyle{plain}

\usepackage{color}

\newdateformat{daymonthyear}{\THEDAY \, \monthname[\THEMONTH] \THEYEAR}
\newdateformat{monthyear}{\monthname[\THEMONTH] \THEYEAR}

\pgfplotsset{compat=1.17}

\begin{document}
\title{Detecting the Unruh Effect via an Engineered Low-Mass Field in a Superconducting Qubit}
\author{Vladimir Toussaint$^{1}$\thanks{{\tt vladimir.toussaint@nottingham.edu.cn}} }
\affil{$^{1}$School of Mathematical Sciences,\\ University of Nottingham Ningbo China,\\
Ningbo 315100, PR China}

 \date{\daymonthyear\today}

\maketitle

\begin{abstract}
Detecting the Unruh effect is a major challenge in fundamental physics. It is known that exciting massive fields with the Unruh thermal bath is heavily suppressed when the field's rest energy is much larger than the acceleration energy scale, $Mc^2 \gg\hbar a/c$. However, the standard literature lacks an explicit quantitative derivation of this suppression. In this work, we first fill this gap by deriving the exponential suppression, $\sim \exp(-\text{constant}\times Mc^2/(\hbar a/c))$, in two different frameworks: a (3+1)-dimensional Unruh-deWitt detector and a (1+1)-dimensional cavity QED setup. This shows the suppression is universal and sets an insurmountable barrier for any detection method that relies on exciting massive fields. For an electron-mass field at achievable accelerations ($a \sim 10^{20}$ m/s$^2$), the suppression exceeds $10^9$ orders of magnitude.
To avoid this suppression, the field's rest energy must be less than or of the order of the acceleration energy scale, $M c^2 \lesssim \hbar a / c$. Achieving this condition, however, requires astronomically high accelerations. For example, detecting the effect for an electron-mass field would require accelerations of  $a \gtrsim 4.6\times 10^{29}$ m/s$^2$, which is far beyond experimental reach.
While using a massless field avoids this suppression, we show the best strategy is not to avoid mass, but to engineer a small effective mass that satisfies the optimal condition $\hbar a / c \gg M_{\text{eff}} c^2$. We propose a concrete implementation using a superconducting circuit with a Josephson persistent-current qubit (the analog of a UdW detector) coupled to a microwave resonator (the analog of the scalar field). For this system, the optimal condition is $2I_p\Delta\Phi \gg \Delta$, where $I_p$ is the persistent current, $\Delta\Phi$ is the magnetic flux swing, and $\Delta$ is the qubit's tunneling energy gap. For a fixed experimental setup $(\lambda, T, \omega_r)$, at the optimal condition, we predict a linear excitation probability $P_{\text{e}}^{\text{signal}}(\delta\omega) \approx \mathcal{S}\delta\omega$. Here, $\delta\omega:=2I_p\Delta\Phi/\hbar$ is the frequency modulation depth and $\mathcal{S}(\lambda,T, \omega_r)$ is a constant that determines the slope. This provides a direct falsifiable prediction for observing the Unruh effect in the optimal regime using a superconducting qubit.
\end{abstract}

\maketitle

\section{Introduction}

The Unruh effect \cite{Unruh:1976db} is a central prediction at the intersection of quantum field theory and general relativity. It states that an observer undergoing constant proper acceleration $a$ will perceive the Minkowski vacuum as a thermal bath of particles at a temperature $T_U = \frac{\hbar a}{2\pi c k_B}$. This effect links acceleration, temperature, and the quantum vacuum \cite{Davies1975, Fulling:1972md, Gibbons:1977mu, Crispino2008}.  Intimately related to the Unruh effect is the Hawking effect, which predicts that black holes emit thermal radiation due to quantum effects near their event horizons \cite{Hawking1975}. Despite their profound theoretical implications, both effects are extremely difficult to observe experimentally. The Unruh effect, in particular, requires accelerations on the order of $2.47 \times 10^{20}$ m/s$^2$ for a 1 K equivalent temperature. This has motivated various strategies to make detection feasible, which can be grouped into two classes.

The first class seeks to observe the Unruh effect in real accelerations but uses more sensitive observables than the direct excitation rate. A prominent example was the investigation of electron storage rings. Seminal work by Bell and Leinaas \cite{Bell:1986ir, Leinaas:2000mh} aimed to interpret the well-known Sokolov-Ternov effect \cite{SokolovTernov1963,DerbenevKondratenko1973,Baier1971}
—where an electron beam reaches an equilibrium polarization of ~92\%—as a signature of the Unruh effect. The idea was that the accelerated electrons would experience a thermal bath, leading to a partially populated excited spin state. However, a detailed analysis revealed a ``null result": the observed 92\% polarization is correctly described by standard QED calculations but cannot be simply attributed to a thermal population at temperature $T_U$. The complications of circular motion (e.g., Thomas precession) mix the signature with other effects, making a direct claim of observation impossible. This motivated alternative approaches, such as the proposal by Chen and Tajima \cite{Chen:1998kp} to use the violent ``snowplow" acceleration of electrons in a laser-driven plasma to directly detect Unruh radiation. 
A more recent proposal by Martin-Martinez et al. \cite{Martin-Martinez:2010gnz} seeks to measure the Berry phase acquired by an accelerated detector, which can be observed at lower accelerations ($ \sim 10^{17}$ m/s$^2$).  Other recent proposals have introduced distinct methods to amplify the faint signal: engineering a resonant cavity to enhance the density of field modes and effectively boost the Unruh temperature for a co-accelerated detector  \cite{Stargen:2021vtg}; using the radiative properties of a rotating atom inside a cavity to probe acceleration-induced modifications of the Wightman function \cite{Lochan:2019osm}; and employing collective superradiance in an atomic array, filtered by a sub-resonant cavity, to extract a time-resolved, orders-of-magnitude amplified burst from the broadband Unruh spectrum \cite{Deswal:2025cjw}.


The second class uses analog quantum simulators, which create an ``effective metric" in a lab system to mimic event horizons. This approach started with Unruh's 1981 proposal that a transsonic fluid flow should emit a thermal spectrum of sound waves, analogous to Hawking radiation \cite{Unruh1981}. This has been realized and extended in several platforms. Proposals based on Bose-Einstein condensates (BECs) study the formation of sonic horizons in supersonic flows \cite{Garay:1999sk}, with recent work demonstrating the quantum simulation of Unruh radiation \cite{Hu:2018psq} and proposing interferometric Unruh detectors \cite{Gooding:2020scc} and Lorentz-invariance-violation-induced nonthermal effects \cite{Tian:2022gfa}. Fiber-optic setups have demonstrated classical effects of an artificial horizon \cite{Philbin:2007ji}. Furthermore, devices like dc-SQUID array transmission lines have been proposed to create an effective metric with a horizon \cite{Nation:2009xb}. Superconducting circuits have also been used to simulate relativistic quantum teleportation \cite{Friis:2012cx} and relativistic motion  \cite{Felicetti:2015kta}, as well as to explore analog cosmological particle generation \cite{Katayama2020} and nonclassical correlations in analog black holes \cite{Katayama2025}. The timelike Unruh effect has been simulated via Berry phase measurements \cite{Quach:2021vzo}  and with superposed trajectories \cite{Cheng2026}, culminating in experimental demonstrations with trapped-ion systems \cite{Luo:2025iur}. The major advantage of these systems is their ability to generate measurable Hawking temperatures and detect correlations between emitted photon pairs—a feat impossible in astrophysics. Moreover, analogue systems are now being used to simulate more than just Hawking’s thermal spectrum — e.g., cosmological expansion, Gibbons–Hawking effect \cite{Fedichev:2003id, Fischer:2004bf}.

In this work, we show a key limitation of both classes of approaches to detecting the Unruh effect when they depend on  massive field excitation and we propose solutions to address it. Specifically, we study two theoretical distinct frameworks: the standard (3+1)-dimensional Unruh-DeWitt (UdW) detector model coupled to a massive scalar field, and a (1+1)-dimensional cavity QED model where an excited confined massless Dirac field decays while exciting a massive unconfined external field. When a massive field is coupled and its rest energy $Mc^2$ greatly exceeds the acceleration energy scale $\hbar a/c$, the response is exponentially suppressed. While this suppression is known in principle \cite{Crispino2008}, its impact on experiments has not been fully appreciated, and the exact scaling has not been derived. We begin by deriving the universal exponential law, $\sim\exp(-\text{constant}\times Mc^2/(\hbar a/c))$, across our two frameworks. Our analysis shows a suppression exceeding $10^9$ orders of magnitude for an electron-mass field at achievable accelerations.

To avoid suppression, the field's rest energy must be less than or of the order of the acceleration energy scale, $M c^2 \lesssim \hbar a / c$ (or $a\gtrsim \frac{Mc^3}{\hbar}$). For an electron-mass field, this requires an astronomical acceleration of $a \gtrsim 4.6\times 10^{29}$ m/s$^2$, which is impractical. While the required acceleration is proportional to the field's mass ($M$), so lighter fields are easier to excite, this isn't practical with real particles. Most analog gravity experiments use massless fields to avoid this suppression, but this creates a new challenge: observing the maximum response requires detectors with impractically small energy gaps. We propose a different strategy: engineering an analog system with a suitably small effective mass, $M_{\text{eff}}$. This lets us satisfy the condition with realistic accelerations, avoiding the suppression. We also show that these systems have a key advantage. For a massive field in the optimal regime $M_{\text{eff}} c^2 \ll \hbar a / c$, by setting the detector gap to the particle creation threshold, $\Delta E\sim M_{\text{eff}} c^2$, we automatically operate at the optimal point $\Delta E \ll \hbar a / c$. This avoids the challenge of fabricating detectors with extremely small gaps ($\Delta E$), which is required for maximizing the response for the massless field case.

Building on this, we propose an experiment using the second class of strategies in a superconducting circuit platform. We use a persistent-current flux qubit coupled to a microwave resonator, where the parameters can be tuned into the optimal regime. We predict a clear linear dependence of the qubit's excitation probability on the applied flux modulation depth, $P_{\text{e}}^{\text{signal}} \approx \mathcal{S}\delta\omega$. This gives a clear path to observing the signature of the Unruh effect in a controlled quantum simulator, overcoming the limitations of both massive and massless field approaches.




\section{\label{sec:FCNon-S}The Fundamental Condition for a Non-Suppressed Effect:}

Observing Unruh-induced effects—whether in the excitation rate of an UdW detector coupled to a massive field, or in a cavity QED model where an excited massless field decays while exciting a massive external field—depends on a simple energy scale comparison. The characteristic energy of the acceleration is set by the Unruh temperature, $k_B T_U = \hbar a/(2\pi c)$. For the effect to be observable, the acceleration energy must be large enough to excite the field, requiring $\hbar a/c \gtrsim Mc^2$. When this condition is violated, $\hbar a/c < Mc^2$, the response is suppressed. For a very massive field, where $Mc^2 \gg \hbar a/c$, this suppression is exponential and severe, making the Unruh-induced effect impossible to detect.

\section{Model 1: Unruh-DeWitt Detector}
We start with the UdW detector model in (3+1)-dimensional Minkowski spacetime to illustrate the exponential suppression.
The UdW model is a point-like two-level system (the detector) with energy levels $|E_0\rangle$ (ground state) and $|E\rangle$ (excited state), separated by an energy gap $\Delta E = E-E_0 >0$. The detector couples locally to a massive Klein-Gordon scalar field $\hat{\phi}(x)$ via its monopole operator $\hat{m}(\tau)$, where $\tau$ is the proper time along its worldline $x^\mu(\tau)$. The interaction is described by the action \cite{Crispino2008}:
\begin{align}\label{UdW:Interaction}
\hat{S}_{I} = \int d\tau\hat{m}(\tau) \hat{\phi}[x(\tau)]\, .
\end{align}
The key parameter is the transition matrix element
$q\equiv \langle E|\hat{m}(0)|E_0\rangle$, which is dimensionless and measures the transition strength.

For such a UdW detector undergoing constant proper acceleration $a$, the excitation rate (probability per unit proper time) for a transition from $|E_0\rangle$ to $|E\rangle$ is given in natural units by [\cite{Crispino2008}, Eq. (3.15)]:
\begin{align}\label{ExRate:MassieField}
R = \frac{|q|^2 e^{-\pi \Delta E/a}}{2\pi^2/a}\int_0^\infty d\mu \mu \left(K_{i\Delta E/a}[\sqrt{\mu^2 + (M/a)^2}] \right)^2 \, .
\end{align}
In the regime $M \gg a$ and  under the threshold condition for excitation, $\Delta E\gtrsim M$, the integral gives a suppression that scales as $\sim \exp(-2M/a)$. This comes from the asymptotic form, $K_{\nu}(z) \sim \sqrt{\pi / (2 z)} \, e^{-z}$, of the modified Bessel function for large arguments, $z\gg 1$, \cite{DLMF}. The explicit Boltzmann factor $e^{-\pi \Delta E/a}\approx e^{-\pi M/a}$ in the prefactor leads to a total suppression of the UdW excitation rate. In SI units, this scaling is
\begin{align}
R\sim \exp(-(\pi +2) M c^2 / (\hbar a/c)) \, .
\end{align}
Thus, for the UdW detector, the excitation rate is exponentially suppressed when $Mc^2\gg \hbar a /c$, which confirms the general barrier from the energy-scale comparison.

\section{Consequence for Fundamental Particles: The Electron Example}
The condition for a non-suppressed Unruh effect, $\hbar a/c \gtrsim Mc^2$, is strongly violated for massive fields like an electron  field. Consider an electron ($M_e c^2 \approx 0.511\, \text{MeV}$) under an extreme achievable acceleration of $a =10^{20} \text{m/s}^2$:
\begin{align}
\frac{\hbar a}{c} &\approx 2.2 \times 10^{-4} \text{eV}, \\
k_B T_U &= \frac{\hbar a}{2\pi c} \approx 3.5 \times 10^{-5} \text{eV}, \\
\frac{M_e c^2}{\hbar a/c} &\approx 2.3 \times 10^9.
\end{align}
This shows that any system with an electron-mass field falls deep into the suppressed regime, $Mc^2 \gg\hbar a/c$, at achievable accelerations.

\section{Model 2: Cavity Setup}
To show that this exponential suppression is universal, we analyze a complementary (1+1)-dimensional model. It has a massless Dirac field $\chi$ confined to a cavity of proper length $l$ with walls at $x_-$ and $x_- + l$. This field couples linearly and locally to an external massive Dirac field $\psi$ of mass $M > 0$. The confinement is enforced by MIT bag boundary conditions \cite{MIT-bag-original} and their probabilistic generalizations \cite{Friis:2011yd, Friis:2013eva}.

The interaction is governed by the Hamiltonian \cite{RefX}:
\begin{equation}
H_{\text{int}} = g \int_{x_-}^{x_- + l} dx \, \left( \bar{\psi} \chi + \bar{\chi} \psi \right),
\end{equation}
where $\bar{\psi}$ and $\bar{\chi}$ are the Dirac adjoints, and $g$ is a small coupling constant with mass dimension $[g] = 1$ (in natural units $\hbar = c = 1$). This ensures $H_{\text{int}}$ has the correct dimension of energy.

This specific coupling is a theoretical model made to study the essential physics. It captures the fundamental energy-scale mismatch that governs the detectability of Unruh-induced enhancements of decay rates from the thermal bath the cavity is uniformly accelerated.

\subsection{The Exponential Suppression in Cavity QED setup}
Our analysis \cite{RefX} shows that for $M c^2 \gg \hbar a / c$, the decay rate of the confined excited field is always dominated by an exponential factor, regardless of cavity size:
\begin{equation}
\Gamma_{\text{acc}}/\Gamma_{\text{in}} \sim \exp(-2 M c^2 / (\hbar a/c))\, ,
\end{equation}
where $\Gamma_{\text{in}}$ is the inertial decay rate.

Although an exact cancellation of exponential terms is possible in a special case with picometer-scale cavities, this fine-tuning is physically impossible and irrelevant for real experiments. For all practical purposes, exponential suppression is the main behavior.

\subsection{Universality}

This comparison shows the exponential scaling and explains the numerical differences between the two models. The UdW model gives a suppression $\sim \exp(-(\pi + 2)M/a)$, while the cavity QED model gives $\sim \exp(-2M/a)$. The extra $\pi$ in the UdW exponent comes from the Boltzmann factor $e^{-\pi \Delta E/a}$, which describes the thermal response of a point-like detector and is not present in the cavity decay rate.

The appearance of this exponential suppression, $\exp(-\text{constant} \times \frac{M}{a})$, in two very different models—a (1+1)-dimensional cavity QED model with Dirac fields and a (3+1)-dimensional UdW model with a Klein-Gordon scalar field—shows that it is not a quirk of a particular model. Instead, it is a universal result of the fundamental energy-scale mismatch, $M \gg a$. The agreement between these models shows that the exponential barrier is a general feature of any attempt to detect Unruh-type effects by exciting massive fields when $M\gg a$ (in natural units).

\section{Fundamental Barrier Across Experimental Approaches}

The exponential suppression, $\sim \exp(-\text{constant}\times C/a)$ with $C = Mc^3/\hbar$, in the regime $Mc^2 \gg\hbar a/c$, is a fundamental barrier. It cannot be overcome by building better accelerators. This limitation applies to all detection methods and field types in both (1+1) and (3+1)-dimensions.

This barrier is impossible to overcome. To see why, consider an electron-mass field ($Mc^2 \sim 0.511 \text{MeV}$). For the cavity QED model, the constant is $C \approx 2.3\times 10^{29}$ m/s$^2$. At a realistic acceleration of $a = 10^{20} \, \text{m/s}^2$, the exponent is $C/a \approx 2.3 \times 10^{9}$, giving a suppression factor of $\exp(-4.6 \times 10^9)$. Increasing the acceleration by a factor $K$ only reduces the exponent to $C/(Ka)$. To get the suppression down to a measurable level ($\sim e^{-1}$), we would need $K \gtrsim 4.6\times 10^{9}$—a completely unrealistic increase. Even with this, the suppression would only be $\sim e^{-1}$. The same conclusion applies to both detection models (the UdW model needs an even larger $K$ due to its larger exponent), showing that the effect is unobservable because of the energy-scale mismatch. This means that to observe the effect, we must use either massless fields or engineered quantum systems that satisfy the condition  $M_{\text{eff}}c^2 \lesssim \hbar a_\text{eff}/c$.

\section{The Path Forward: Quantum Simulation}

The solution to this fundamental constraint is not to achieve higher accelerations, but to engineer systems where the effective mass energy satisfies $M_{\text{eff}}c^2 \lesssim \hbar a_\text{eff}/c$. The optimal regime is when $M_{\text{eff}}c^2 \ll \hbar a_\text{eff}/c$ (i.e., the mass energy is much less than the acceleration energy scale). In this regime, the exponential suppression is negligible and the response is easily measurable.

This approach is realized in quantum simulators like superconducting circuits and optomechanical systems \cite{Friis:2012cx, Aspelmeyer:2013lha}. In these platforms, we can simulate high effective accelerations of $a_\text{eff}\sim 10^{20} \text{m/s}^2$ by modulating parameters in time. We can also design the system to have a small effective mass, with $M_{\text{eff}}c^2 \sim 10^{-5} -10^{-4} \text{eV}$. With $\hbar a_\text{eff}/c \sim 10^{-4} \text{eV}$, these parameters easily satisfy the condition for observability, $M_{\text{eff}}c^2 \lesssim \hbar a_\text{eff}/c$. For the smallest masses in this range ($M_{\text{eff}}c^2 \sim 10^{-5} \text{eV}$), they even satisfy the optimal condition, $M_{\text{eff}}c^2 \ll \hbar a_\text{eff}/c$.

This quantum simulation approach not only solves the fundamental barrier but also offers an unexpected benefit over massless field detection schemes, as we now demonstrate.

\section{Advantage of Engineered Massive Fields in the Observable Regime}

We find a surprising advantage: in the engineered regime $M_{\text{eff}}c^2 \ll \hbar a/c$, a massive field can be easier to observe than a massless field under the same conditions. This happens because the effective rest energy, $M_{\text{eff}}c^2$, naturally sets the detector's energy gap to $\Delta E\sim M_{\text{eff}}c^2$. This ensures the system operates in the regime, $\Delta E\ll \hbar a/c$, without any fine-tuning. These arguments are general and apply to any UdW-type detector-field setup, indicating that engineering a small effective mass is a universal strategy to overcome the exponential suppression.

To see why, consider a massless field. The excitation rate for an Unruh-deWitt detector is \cite{Crispino2008}:
\begin{align}
R_{\text{massless}} = \frac{|q|^2 \Delta E}{2\pi\hbar} \frac{1}{e^{2\pi c \Delta E/(\hbar a)}-1}\, .
\end{align}
The rate is maximized when $\Delta E \ll \hbar a/c$, giving
\begin{align}\label{Engi:Rate}
R_{\text{massless}} \approx \frac{|q|^2 a}{4\pi^2 c}\, .
\end{align}

But to achieve $\Delta E\ll \hbar a/c$, we would need to engineer a detector with an extremely small energy gap. This is experimentally challenging because low-energy signals are easily drowned out by thermal noise, and there are technical limits on how small a gap we can make.

In contrast, for a massive field with $M_{\text{eff}}c^2 \ll \hbar a/c$, the excitation threshold condition, $\Delta E \gtrsim M_{\text{eff}}c^2$, gives a natural solution. If we set the gap to $\Delta E\sim M_{\text{eff}}c^2$ (the minimum needed to create a particle), we automatically satisfy $\Delta E \ll \hbar a/c$. The correct leading-order approximation for the excitation rate from Eq. \eqref{ExRate:MassieField} then becomes (in SI units):
\begin{align}
R_{\text{eff, massive}} \approx \frac{|q|^2 a}{4\pi^2 c}\, ,
\end{align}
which is the same as the massless case at its optimum.

The main advantage is that the mass provides a natural scale that ensures optimal conditions without fine-tuning the detector. Both cases give the same maximum rate when optimized, but the massive field case gets there automatically by setting $\Delta E\sim M_{\text{eff}}c^2$. The massless case, however, requires delicate engineering of a very small $\Delta E$. This difference shows that engineered quantum systems operating in the $M_{\text{eff}}c^2\ll \hbar a/c$ regime provide a more practical and accessible way to probe Unruh-type effects.

\section{Experimental Proposal: Analog Simulation of Unruh-deWitt Detector Response}

We propose an experiment to simulate an Unruh-deWitt detector using a superconducting quantum circuit. The goal is to verify a key prediction of Unruh theory: that the excitation rate $R_{\text{eff, massive}}$ is linear in the acceleration $a$, following $R_{\text{eff, massive}} \approx \frac{|q|^2 a}{4\pi^2 c}$, in the optimal regime, $M_{\text{eff}} c^2 \ll \hbar a / c$. We will test this by observing a linear relationship between the qubit's excitation probability and the frequency modulation depth $\delta\omega$. This is an analog simulation, not a direct recreation of an accelerating object.

It is important to distinguish this proposed analogue simulation of the Unruh effect from the dynamical Casimir effect (DCE). In the Unruh effect, the detector's internal energy gap is modulated (simulating its proper time evolution in an accelerated frame), while the field's boundary conditions remain static. In contrast, DCE arises from time-dependent boundary conditions of the field itself. This distinction has been clarified in analogue gravity implementations with cold atoms \cite{Fedichev:2003bv} and superconducting circuits \cite{Tian:2017wij}. Our setup, where the flux modulation affects the qubit's gap while the resonator remains fixed, squarely falls into the Unruh effect simulation.
 
Furthermore, 
while previous theoretical proposals using superconducting circuits represent important theoretical proposals for studying relativistic effects using superconducting circuits,  they differ significantly in scope, physical setup, and objectives from our proposal. For instance, Friis \textit{et al.} \cite{Friis:2012cx} studied how non-uniform cavity motion degrades quantum teleportation fidelity via particle creation, proposing a SQUID-terminated waveguide experiment where the signature is a fidelity reduction corrected by local phases. Felicetti \textit{et al.} \cite{Felicetti:2015kta} simulated qubit motion via coupling modulation, observing acceleration radiation and anti-Jaynes–Cummings dynamics, with signatures in photon number and qubit polarization. In contrast, our proposal focuses on the massive field case and explicitly engineering the optimal condition $M_{\text{eff}} c^2 \ll \hbar a_{\text{eff}}/c$ to overcome the exponential suppression. Moreover, we provide a concrete prediction for the linear response of the excitation probability with modulation depth, which serves as a clear experimental signature.

We use a superconducting persistent-current flux qubit \cite{Orlando:1999xxc, Moodij:1999,Schumacher:1995nrx, NakamuraPashkinTsai1999, Krantz:2019jkw, DevoretWallraffMartinis2004} as the analog detector. The qubit is coupled to a linear microwave resonator \cite{Wendin:2017ymy, Blais2005, Koch2007, CastellanosBeltran2008, Pozar2011}, which represents a one-dimensional scalar field. We simulate acceleration by applying a large-amplitude, high-frequency flux bias to the qubit \cite{ClarkeWilhelm2008}. By measuring the excitation probability as a function of the control parameter for $a$, we test the predicted linear relationship.

The qubit's dynamics are modeled by a two-level Hamiltonian in the persistent-current basis \cite{Moodij:1999, Paauw2009}:
\begin{align}
 H_q= -\frac{1}{2}\left [\epsilon(\Phi_\text{ext})\sigma_z + \Delta \sigma_x\right]
\end{align}
where $\sigma_{x,z}$ are Pauli matrices. The parameter $\epsilon(\Phi_\text{ext})= 2I_p\left(\Phi_{\text{ext}}-\frac{\Phi_0}{2} \right)$ is the flux-tunable energy bias, with $I_p$ the persistent current and $\Phi_0 =h/(2e)$ the flux quantum. The parameter $\Delta$ is the tunneling energy. The qubit transition frequency is $\omega_q=  \frac{1}{\hbar}\sqrt{\epsilon^2 + \Delta^2}$.

The scalar field is modeled as a high-Q superconducting microwave resonator with Hamiltonian $\hat{H}_r = \hbar \omega_r (\hat{a}^\dagger \hat{a} +\frac{1}{2})$, where $\hat{a} (\hat{a}^\dagger)$  is the annihilation (creation) operator for the fundamental resonant mode of frequency $\omega_r$. Thus, we treat the qubit's interaction as coming mainly from a single bosonic mode with energy $\hbar \omega_r$. This single-mode approximation is standard in circuit QED and is valid here because all other modes are far detuned—much farther from the main frequency than the qubit-resonator coupling strength $\lambda$ (see for e.g. \cite{Wendin:2017ymy, Blais2005}). The resonator's fundamental frequency $\omega_r$ introduces a gap in the excitation spectrum. This gap acts as an effective rest energy for the resonator photons, $M_{\text{eff}} c^2 = \hbar \omega_r$, analogous to the mass term in the Klein-Gordon equation. Thus, the resonator mimics a massive scalar field with effective mass $M_{\text{eff}}$. While the microwave resonator supports a massless field mode, in our analogue model we treat its energy gap as an effective rest mass  $M_{\text{eff}}$ to study the suppression behavior predicted for massive fields. This is a controlled idealization to explore the $M\gg a$ regime in a system where true mass is absent. Thereby, 
   we define the engineered effective mass as $M_{\text{eff}}c^2 := \hbar \omega_r$. To operate in the optimal regime, we require $\omega_r\ll a_{\text{eff}}/c$.

   The qubit and resonator are coupled via the full quantum Rabi interaction Hamiltonian \cite{Rabi:1936lvg, JaynesCummings1993, Braak:2011tmj}:
    \begin{align} \label{JC:Hamiltonian}
   H_{\text{int}} = \hbar\lambda\,(\sigma^+ + \sigma^-)(a + a^\dagger)\, ,
    \end{align} 
    with a coupling strength $\lambda/(2\pi) \approx 10-50 \text{MHz}$. In this analogue, the qubit's Pauli operator $\sigma^+$ ($\sigma^-$) corresponds to the detector's monopole moment raising (lowering) operator, while the resonator's annihilation (creation) operator $a$ ($a^\dagger$) corresponds to the field operator evaluated at the detector's position. The full Rabi model includes both the energy-conserving Jaynes-Cummings terms $(\sigma^+ a + \sigma^- a^\dagger)$ and the counter-rotating terms $(\sigma^+ a^\dagger + \sigma a)$. The counter-rotating term $\sigma^+ a^\dagger$ is essential for enabling transitions from the joint ground state $|g,0\rangle$ to the excited state $|e,1\rangle$, as required by the Unruh effect simulation. As shown in Appendix~\ref{app:full_rabi_excitation}, the combination of this term with the time-dependent modulation $\omega_q(t)$ makes such transitions possible.

To simulate acceleration, we modulate the external flux bias around the degeneracy point:
\begin{align}\label{Modulation}
\Phi_{\text{ext}}(t) = \Phi_0/2 + \Delta\Phi \cos(\Omega t)\, ,
\end{align}
where $\Delta\Phi$ is the flux swing and $\Omega$ is the modulation frequency. At the degeneracy point ($\Phi_{\text{ext}}^{(DC)}=\Phi_0/2$), the energy bias is zero and the gap is $\hbar \omega_q = \Delta$. The time-dependent flux modulates the qubit's transition frequency:
\begin{align}
\omega_q(t) = \frac{\Delta}{\hbar} \sqrt{ \left(\frac{2I_p \Delta\Phi}{\Delta} \cos(\Omega t)\right)^2 + 1 }\, .
\end{align}
This modulated qubit's transition frequency $\omega_q(t)$ mimics the time dilation experienced by an accelerated detector, effectively simulating its proper time evolution in a relativistic frame. 
 To quantify the strength of the simulated acceleration in our analogue model, we introduce an effective Lorentz factor $\gamma(t)$ and an effective acceleration $a_\text{eff}$ by analogy with Unruh-deWitt theory. We define an effective Lorentz factor $\gamma(t)=\hbar \omega_q(t)/\Delta$ and a peak effective acceleration $a_\text{eff, max}$ from the maximum rate of change of $\gamma(t)$.
By setting the modulation frequency to match the qubit's gap, $\Delta/\hbar =\Omega$, and under the large modulation condition $2I_p \Delta\Phi\gg \Delta$, we get:
\begin{align}\label{OptimalAnalog:Regime}
a_\text{eff, max} \approx c\frac{2I_p \Delta\Phi}{\hbar} = c \delta\omega \, ,
\end{align}
where $\delta\omega := \frac{2I_p \Delta\Phi}{\hbar}$ defined the frequency modulation depth. The optimal regime is when the resonator frequency is much smaller than the effective Unruh frequency, $\omega_r \ll a_{\text{eff, max}}/c$, or equivalently, $\omega_r\ll \delta\omega$. For a high-Q resonator frequency of $\omega_r/(2\pi) = 5\, \text{GHz}$ and a modulation depth of $\delta\omega/(2\pi) = 50 \, \text{GHz}$, the optimal condition is satisfied, yielding a peak effective acceleration of $a_{\text{eff, max}}\approx  10^{20} \text{m/s}^2$. This corresponds to an effective Unruh temperature of $T_U = \hbar a_{\text{eff, max}}/(2\pi c k_B) \approx 3.5 \times 10^{-5} \, \text{eV}/k_B \approx 0.4 \, \text{K}$, which is much higher than the typical operating temperature of a dilution refrigerator ($\sim 10 \, \text{mK}$), ensuring that the signal is not masked by thermal noise.

To ensure the validity of our model, we must operate within the regime where the superconducting flux qubit remains a well-defined two-level system. It is established that a flux qubit is in its operational regime when the penetrating flux is close to half a superconducting flux quantum, $\Phi_0/2$ (mod $\Phi_0$), where the two lowest energy eigenstates are well separated from higher states \cite{Paauw2009}. We bias the qubit at the degeneracy point and apply a small-amplitude flux modulation Eq.~\eqref{Modulation}. In the optimal regime, the frequency modulation depth $\delta \omega$ is related to the flux swing $ \Delta\Phi$ by the persistent current $I_p$ of the qubit junction via $\delta\omega = 2I_p \Delta\Phi/\hbar$ (Eq.~\eqref{OptimalAnalog:Regime}). For a typical persistent current of
$I_p\approx 1\, \mu\text{A}$, achieving the desired modulation depth of $\delta\omega/(2\pi) \sim 50 \, \text{GHz}$ requires a flux swing of only $\Delta\Phi\approx 0.008\Phi_0$. This corresponds to a magnetic flux varying between  $\Phi_{\text{min}}\approx 0.492\Phi_0$ and $\Phi_{\text{max}}\approx 0.508\Phi_0$, which is within 0.8\% of the degeneracy point. Thus, the qubit model remains valid. The ``large modulation condition'' refers to the energetic drive (50 GHz) being large compared to the qubit gap (~5 GHz), not a large physical displacement.

With the operational regime established, we now describe the measurement protocol.
We measure the excitation probability $P_{e}$ of the qubit after a fixed exposure time $T$. In our analogue model, this signal is interpreted as being generated by the effective massive rate, such that $P_{\text{e}}= R_{\text{eff, massive}}T=  \left(\frac{|q|^2 a_{\text{eff, max}}}{4\pi^2 c}\right) T$. In this context, $|q|^2$ is not the UdW matrix element; it is related to the qubit-resonator coupling strength, $\lambda$ by  $|q|^2 = 4\pi^2 \mathcal{K}|\lambda|^2$, where $\mathcal{K}$ is a constant that depends on the interaction details with dimensions of time squared.

We prepare the system in the joint ground state (qubit in $|g\rangle$, resonator in $|0\rangle_{\text{ph}}$). This is like a detector at rest in the vacuum. We then apply the time-dependent  flux-bias pulse $\Phi_\text{ext}(t)$ for a duration $T$ (the ``flight time''). As shown in Appendix~\ref{app:full_rabi_excitation}, this pulse modulates the qubit's transition frequency $\omega_q(t)$, which dynamically tunes the system into resonance with a process that excites both the qubit and the resonator from the vacuum. This explicit time-dependence—the simulated ``motion"—provides the energy required for this otherwise forbidden transition. This process directly simulates the core excitation mechanism of the Unruh effect, where a detector's motion leads to a perception of particles from the vacuum.
 After the pulse, we switch off the modulation and perform a dispersive readout to measure $P_{\text{e}}(\delta\omega; T)$.  To isolate the signal, we subtract the background excitation from a control experiment with no modulation ($\Delta \Phi =0$):
\begin{align}
  P_{\text{e}}^{\text{signal}}(\delta\omega; T)       = P_{\text{e}}(\delta\omega; T)- P_{\text{e}}^{\text{idle}}(\delta\omega; T)\, .
\end{align}

Our analysis predicts the excitation probability is:
\begin{align}
  P_{\text{e}}^{\text{signal}}(\delta\omega; T)       \approx \left(\frac{\mathcal{K}|\lambda|^2 a_{\text{eff, max}}}{ c} \right)T\,.
\end{align}
Using $a_{\text{eff, max}}\approx c\delta\omega$, we get
\begin{align}
  P_{\text{e}}^{\text{signal}}(\delta\omega; T)       \approx (\mathcal{K}|\lambda|^2 T) \delta\omega\, , \quad \delta\omega=\frac{2I_p \Delta\Phi}{\hbar}  \, .
\end{align}
This linear scaling emerges because the large modulation depth activates many sidebands, and the total transition rate sums coherently over these sidebands, yielding a result proportional to $\delta\omega$ (see Appendix~\ref{app:full_rabi_excitation} for the detailed derivation).
This is our main prediction: for a fixed setup $(\lambda, T, \omega_r)$, the excitation probability is linear with $\delta \omega$. A plot of $P_{\text{e}}^{\text{signal}}$ versus $\delta\omega$ will be a straight line through the origin with slope $S =\mathcal{K}|\lambda|^2 T$, confirming the core linear response predicted by the Unruh effect in the optimal regime where $\omega_r \ll \delta \omega$. 

 We can also write this linear relation as:
\begin{align}
 P_{\text{e}}^{\text{signal}}(\Delta\Phi; T)      \approx \mathcal{C}_\text{sys} \Delta\Phi\, , \quad \mathcal{C}_\text{sys} =  \frac{2I_p \mathcal{K}|\lambda|^2 T }{\hbar}  \, .
\end{align}
This shows that the flux swing $\Delta\Phi$ controls the signal linearly. We also predict that the slope scales inversely with the square of the resonator frequency, $S\propto 1/\omega_r^2$, leading to:
\begin{align}
P_{\text{e}}^{\text{signal}}(\Delta\Phi;T) \approx \mathcal{C}'\frac{|\lambda|^2 T}{\omega_r^2} \delta\omega\, ,
\end{align}
where  $ \mathcal{C}'=\omega_r^2\mathcal{K}$ is a dimensionless constant.

The main difficulty facing this experiment is maintaining qubit coherence during the pulse of duration $T$. The qubit must stay in a pure state long enough to accumulate a measurable signal. This requires operating at millikelvin temperatures to suppress thermal noise and using a high-coherence qubit, such as the persistent-current qubit we propose.

A rough calculation shows both the promise and the challenge. For representative parameters ($\lambda/2\pi = 30$ MHz, $\omega_r/2\pi = 5$ GHz, $\delta\omega/2\pi = 50$ GHz, $T =2$ ns), the predicted excitation probability is proportional to:
\begin{align}
P_{\text{e}}^{\text{signal}}(\delta\omega; T) \approx \mathcal{C}'\frac{|\lambda|^2 T}{\omega_r^2} \delta\omega\approx 0.0226\times \mathcal{C}' \, ,
\end{align}
where $\mathcal{C}'$ is a dimensionless constant. For the probability to be physically plausible ($\leq 1$), 
$\mathcal{C}'$ must be less than approximately 44. Therefore, this yields a signal that's potentially measurable. The main challenge is maintaining qubit coherence for the duration $T$ in order to carry out the measurement.

The experiment relies on three key assumptions. First, we operate in the resonant regime, where the modulation frequency $\Omega$ equals the qubit's gap at the degeneracy point, $\Omega =\Delta/\hbar$. Second, we assume a `large modulation condition', $2I_p\Delta\Phi \gg \Delta$, which allows us to derive a simple expression for the effective acceleration \eqref{OptimalAnalog:Regime}. Third, we require the system to be in the optimal regime, $\omega_r\ll \delta\omega$. Our proposed parameters are chosen to satisfy all three conditions.



\section{Conclusion}

Our analysis shows a basic energy-scale rule for observing Unruh effects with massive fields. Exponential suppression happens when $M c^2 \gg \hbar a / c$. To see the effect, we need $M c^2 \lesssim \hbar a / c$  (or $a\gtrsim \frac{Mc^3}{\hbar}$). However, this condition requires astronomical acceleration far beyond experimental reach in the foreseeable future. Thus, building better accelerators won't solve this problem for massive fields.
 We see this exponential suppression, $\sim \exp(-\text{constant} \times Mc^2/ (\hbar a/c))$,  in both the (1+1)-dimensional cavity QED and the (3+1)-dimensional Unruh-DeWitt detector models. This explains why detection approaches that rely on exciting massive fields are extremely challenging, as the suppression factor is astronomically large for any realistic experiment (e.g., for a field of electron mass, this requires an acceleration $a \gtrsim 4.6\times 10^{29}$ m/s$^2$).

However, we also show that we can bypass this barrier by engineering systems where the effective mass satisfies $M_{\text{eff}} \lesssim \hbar a/c$. We found a key advantage for engineered quantum systems. For a UdW detector coupled to a massive scalar field, the optimal condition $M_{\text{eff}} \ll \hbar a/c$ automatically sets the detector gap to the optimal point, $\Delta E\sim M_{\text{eff}} \ll \hbar a/c$, without any fine-tuning. This is a big advantage over traditional methods that use massless fields, where low-energy signals are hard to detect and technical limits prevent making the gap very small.

We propose a clear experimental protocol to detect the Unruh effect using a superconducting qubit. We show that if we set up the system so the flux drive energy is much greater than the qubit's minimum gap ($2I_p\Delta\Phi \gg \Delta$), we enter the optimal regime where $\omega_r\ll \delta\omega$ (with $\delta\omega=\frac{2I_p \Delta\Phi}{\hbar}$). We predict a linear relation, $P_{\text{e}}^{\text{signal}}(\delta\omega; T) \approx \mathcal{S}\delta\omega$). Specifically, the model makes a clear falsifiable prediction: the excitation probability should be linear in the modulation depth $\delta\omega$. The slope is predicted to be $\mathcal{S} = \mathcal{K}|\lambda|^2 T$ (with $\mathcal{K}$ having dimensions of time squared), which can also be written as $\mathcal{S}=\mathcal{C}'|\lambda|^2 T/\omega_r$ where $\mathcal{C}'=\omega_r^2\mathcal{K}$ a dimensionless constant. Alternatively, the relation can be written as $P_{\text{e}}^{\text{signal}}(\Delta\Phi; T) \approx \mathcal{C}_\text{sys} \Delta\Phi$, with $\mathcal{C}_\text{sys} = \frac{2I_p \mathcal{K}|\lambda|^2 T }{\hbar}$. This shows that the signal response is linear with the flux swing $\Delta\Phi$, in the optimal regime.

A key feature of our model is the effective constant $\mathcal{K}$, which contains the details of the qubit-field interaction. As derived in the continuum model in 
Appendix~\ref{app:full_rabi_excitation}, $\mathcal{K}$
depends on the specific forms of the density of states $\rho(\omega)$ and the coupling function $\lambda(\omega)$, and is therefore sensitive to the continuum properties of the resonator. Our calculation does not determine its exact value, as this would require a more detailed theoretical analysis of the specific experimental setup. Looking further ahead, our work opens two clear directions for advancing the simulation. First, a primary future task is to determine the value of
the effective constant $\mathcal{K}$ by fitting experimental data, as its value is set by the specific resonator properties. Second, extending the framework to include a time-dependent coupling $\lambda(t)$ would enable simulations of non-uniform acceleration and probe particle production in a broader range of effective spacetimes \cite{Fedichev:2003dj}.

\section*{Acknowledgments}

We would like to thank Jorma Louko and Wan Mohamad Husni Wan Mokhtar for useful discussions.

\appendix
\section{Excitation from the Ground State with Periodic Frequency Modulation}
\label{app:full_rabi_excitation}

This appendix provides a detailed account of the excitation mechanism in the proposed analog Unruh simulation. We begin with the exact physical model, discuss the mathematical challenges of a direct derivation, and justify the use of an effective model that yields a clear and testable prediction.

\subsection{The Exact Model: Large-Modulation Limit and Non-Sinusoidal Frequency Modulation}
The system is described by the full Rabi Hamiltonian \cite{Rabi:1936lvg, JaynesCummings1993, Braak:2011tmj}, where a superconducting qubit is coupled to a microwave resonator. The qubit's transition frequency is modulated periodically by an external flux bias, leading to the Hamiltonian:
\begin{equation}
H(t) = \frac{\hbar}{2}\omega_q(t)\sigma_z + \hbar\omega_r a^\dagger a + \hbar\lambda\,(\sigma^+ + \sigma^-)(a + a^\dagger).
\label{eq:exact_ham_appendix}
\end{equation}
Our analysis on the transition probability focuses exclusively on the $\sigma^+ a^\dagger$ term in the full interaction Hamiltonian, 
 as it is the only term that couples the initial state $|g, 0\rangle$ to the target excited state $|e, 1\rangle$, mirroring the absorption process in the Unruh-deWitt model. The other terms are neglected: the $\sigma^- a$ term cannot act on the ground state, and the remaining terms either describe emission processes or couple to different state manifolds.

The exact form of the modulated frequency, derived from the flux qubit's potential energy, is:
\begin{equation}
\omega_q(t) = \frac{\Delta}{\hbar} \sqrt{1 + \left(\frac{\delta\omega}{\Omega}\right)^2 \cos^2(\Omega t)}.
\label{eq:exact_omega_appen}
\end{equation}
Here, $\Delta$ is the bare qubit gap, $\Omega$ is the modulation frequency, and we define the frequency modulation depth as $\delta\omega := \frac{2I_p \Delta\Phi}{\hbar}$, where $I_p$ is the persistent current and $\Delta\Phi$ is the flux swing. We set the modulation frequency to the bare qubit gap, $\Omega = \Delta/\hbar$.

Under the large
modulation condition, we have: $\delta\omega \gg \Omega$ (or $2I_p \Delta\Phi \gg \Delta$). In this limit, Eq. (\ref{eq:exact_omega_appen}) simplifies to:
\begin{align}
\omega_q(t) &\approx \delta\omega \, |\cos(\Omega t)|.
\end{align}
This results in a rectified sine wave, a non-sinusoidal waveform that is analytically complex, with a Fourier series containing a DC offset and an infinite number of harmonics at multiples of $2\Omega$.

Attempting a direct, rigorous derivation of the transition rate from the exact Hamiltonian is mathematically formidable. The transition amplitude involves an integral of the form:
\begin{equation}
c_{e,1}(t) \propto \int_0^t \exp\left[ i \int_0^{t'} \delta\omega |\cos(\Omega t'')| dt'' \right] dt'.
\end{equation}
The integral $\int |\cos(\Omega t'')| dt''$ does not have a closed-form solution in terms of elementary functions; it is expressed using elliptic integrals. The resulting expression for the amplitude $c_{e,1}(t)$ is an implicit and complicated function of time, making it practically impossible to extract a simple scaling law for the transition rate $\Gamma$.

\subsection{The Effective Model: A Physically Motivated Simplification}
To obtain a clear analytical prediction, we analyze an effective model that captures the dominant physics but is analytically tractable. We replace the complex, non-sinusoidal drive with a purely sinusoidal one of the same amplitude:
\begin{equation}
\omega_q(t) \rightarrow \omega_{q0} + \delta\omega \cos(\Omega t).
\label{eq:effective_modul}
\end{equation}
This leads to the simplified Hamiltonian:
\begin{align}
H(t) &= H_0(t) + H_{\text{int}}, \label{eq:effective_H} \\
H_0(t) &= \frac{\hbar}{2}\bigl[\omega_{q0} + \delta\omega\cos(\Omega t)\bigr]\sigma_z + \hbar\omega_r a^\dagger a, \label{eq:effective_H0} \\
H_{\text{int}} &= \hbar\lambda\,(\sigma^+ + \sigma^-)(a + a^\dagger). \label{eq:effective_Hint}
\end{align}

The use of this effective model is justified. It shares the same key control parameter  $\delta\omega$ as the exact model and correctly describes the core physical mechanism: a dynamical resonance via the $\sigma^+ a^\dagger$. The model is consistent with the acceleration analogy ($a_{\text{eff, max}}\approx c\delta\omega$) and predicts a comparable bandwidth of sidebands for both drives, with the number of contributing channels proportional to $\delta\omega$. Thus, the effective model is physically well-founded.

\subsection{Transition Amplitude for a Single Mode}

We first consider a single resonator mode to illustrate the mechanism. Moving to the interaction picture with respect to the time‑independent parts and treating the modulation as a perturbation, the relevant term coupling $|g,0\rangle$ to $|e,1\rangle$ is
\begin{align}\label{SingleMode:InteractionPicHam}
\tilde{V}(t) = \hbar\lambda\,\sigma^+ a^\dagger\,
               \exp\!\Bigl[i(\omega_{q0} + \omega_r)t + i\frac{\delta\omega}{\Omega}\sin(\Omega t)\Bigr]
\end{align}

Using the Jacobi‑Anger expansion \cite{AbramowitzStegun1965}, $e^{i x\sin(\Omega t)} = \sum_{n=-\infty}^{\infty} J_n(x) e^{in\Omega t}$ with $x = \delta\omega/\Omega$, we obtain
\begin{align}
\tilde{V}(t) = \hbar\lambda \sum_{n=-\infty}^{\infty}
               J_n\!\Bigl(\frac{\delta\omega}{\Omega}\Bigr)\,
               \sigma^+ a^\dagger\,
               e^{i(\omega_{q0}+\omega_r+n\Omega)t} + \text{H.c.}
\end{align}

The first‑order transition amplitude from $|g,0\rangle$ to $|e,1\rangle$ is
\begin{align}
c_{e,1}(t) = -i\lambda \sum_{n} J_n\!\Bigl(\frac{\delta\omega}{\Omega}\Bigr)
             \int_0^t e^{i(\omega_{q0}+\omega_r+n\Omega)t'} dt'.
\end{align}

If the modulation frequency $\Omega$ is chosen such that the resonance condition $\omega_{q0}+\omega_r+n\Omega = 0$ is satisfied for some integer $n$, the corresponding term grows linearly in time. For example, choosing a modulation frequency $\Omega$ that satisfies the resonance condition for $n=-1$ (i.e., $\Omega = \omega_{q0}+\omega_r$) leads to
\begin{align}
c_{e,1}(t) \approx i\lambda J_{1}\!\Bigl(\frac{\delta\omega}{\Omega}\Bigr) t,
\qquad
P_e(t) \approx \lambda^2 J_{1}^2\!\Bigl(\frac{\delta\omega}{\Omega}\Bigr) t^2.
\end{align}

This demonstrates unequivocally that excitation from the ground state is possible when the full Rabi interaction is combined with time‑dependent frequency modulation.

\subsection{Extension to a Continuum and Scaling Argument}
\label{app:continuum}

This section presents a theoretical analysis of a generalized model where the qubit is coupled to a continuum of bosonic modes, in order to derive the predicted scaling law. The actual experiment uses a single-mode approximation, focusing on the resonator's primary mode $\omega_r$.

In this model, the total Hamiltonian is
\begin{align}\label{Effec:Ham}
H(t) &= \frac{\hbar}{2}\omega_q(t)\sigma_z
      + \sum_k \hbar\omega_k a_k^\dagger a_k
      \notag\\   &+ \hbar\sum_k \lambda_k\,(\sigma^+ + \sigma^-)(a_k + a_k^\dagger),
\end{align}
where $k$ labels the modes of the continuum.

The interaction‑picture Hamiltonian for the $\sigma^+ a_k^\dagger$ terms is
\begin{align}
\tilde{V}(t) = \hbar\sum_k \lambda_k \sum_{n} J_n\!\Bigl(\frac{\delta\omega}{\Omega}\Bigr)\,
               \sigma^+ a_k^\dagger\,
               e^{i(\omega_{q0}+\omega_k+n\Omega)t} + \text{H.c.}
\end{align}
Using Fermi's golden rule, the total transition rate is
\begin{align}
\Gamma = 2\pi \sum_{n} J_n^2\!\Bigl(\frac{\delta\omega}{\Omega}\Bigr)\,
        \sum_k |\lambda_k|^2\,
        \delta(\omega_{q0} + \omega_k + n\Omega). 
\end{align}
Converting the sum to an integral over a density of states $\rho(\omega)$ gives
\begin{align}\label{eq:Gamma}
\Gamma = 2\pi \sum_{n} J_n^2\!\Bigl(\frac{\delta\omega}{\Omega}\Bigr)\,
        |\lambda(\omega_n)|^2 \rho(\omega_n),
\qquad
\omega_n = -\omega_{q0} - n\Omega.
\end{align}

The continuum model is characterized by a density of states $\rho(\omega)$ and a coupling function $\lambda(\omega)$, with a characteristic frequency $\omega_c$ (e.g., the frequency of the lowest mode). The optimal regime requires the modulation to be fast compared to this scale, $\delta\omega\gg \omega_c$. For our experiment, the continuum is formed by the resonator's modes, so the characteristic frequency is the fundamental mode's frequency, $\omega_c\approx \omega_r$.  

With $\Omega \sim \omega_c$ and $\delta\omega\gg \omega_c$, the argument of the Bessel functions, $x=\delta\omega/\Omega$, is large. In this limit, $J_n(x)$ is significant for a wide range of sideband orders $|n| \lesssim x$. The number of effectively contributing sidebands therefore scales as
$N_{\text{eff}} \propto \frac{\delta\omega}{\Omega}$.
To estimate the total rate, we must evaluate the sum in \eqref{eq:Gamma}, which depends on the specific forms of $\rho(\omega)$ and $\lambda(\omega)$. If we make the simplifying assumption that the continuum properties, encapsulated in the product $|\lambda(\omega)|^2\rho(\omega)$, vary slowly over the frequency window spanned by these sidebands (which has width $\propto N_{\text{eff}}\Omega \propto \delta\omega$), then each term in the sum \eqref{eq:Gamma} contributes roughly equally. Under this assumption, the total rate scales with the number of contributing sidebands:
\begin{align}
\Gamma \propto N_{\text{eff}} \propto \frac{\delta\omega}{\Omega} \propto \delta\omega.
\end{align}
For a fixed interaction time $T$, the excitation probability scales as $P_e = \Gamma T \propto \delta\omega T$. This linear dependence is the central prediction of the analog simulation, mirroring the Unruh-DeWitt model where the excitation rate is proportional to acceleration. The experiment is designed to test this prediction.



\begin{thebibliography}{9}



\bibitem{Unruh:1976db}
W.~G.~Unruh,
``Notes on black hole evaporation,''
Phys. Rev. D \textbf{14}, 870 (1976).

\bibitem{Davies1975}
P.~C.~W.~Davies,
``Scalar particle production in Schwarzschild and Rindler metrics,''
J. Phys. A \textbf{8}, 609-616 (1975).

\bibitem{Fulling:1972md}
S.~A.~Fulling,
``Nonuniqueness of canonical field quantization in Riemannian space-time,''
Phys. Rev. D \textbf{7}, 2850-2862 (1973).

\bibitem{Gibbons:1977mu}
G.~W.~Gibbons and S.~W.~Hawking,
``Cosmological event horizons, thermodynamics, and particle creation,''
Phys. Rev. D \textbf{15}, 2738–2751 (1977).

\bibitem{Crispino2008}
L.~C.~B.~Crispino, A.~Higuchi and G.~E.~A.~Matsas,
``The Unruh effect and its applications,''
Rev. Mod. Phys. \textbf{80}, 787-838 (2008).

\bibitem{Hawking1975}
S.~W.~Hawking,
``Particle Creation by Black Holes,''
Commun. Math. Phys. \textbf{43}, 199-220 (1975)
[erratum: Commun. Math. Phys. \textbf{46}, 206 (1976)].




\bibitem{Bell:1986ir}
J.~S.~Bell and J.~M.~Leinaas,
``The Unruh Effect and Quantum Fluctuations of Electrons in Storage Rings,''
Nucl. Phys. B \textbf{284}, 488 (1987).


\bibitem{Leinaas:2000mh}
J.~M.~Leinaas,
``Unruh effect in storage rings,''
[arXiv:hep-th/0101054 [hep-th]].



\bibitem{SokolovTernov1963}
A.~A. Sokolov and I.~M. Ternov, 
``On polarization and spin effects in the theory of synchrotron radiation'', 
Dokl. Akad. Nauk SSSR \textbf{153}, 1052 (1963) 
[Sov. Phys. Dokl. \textbf{8}, 1203 (1964)].


\bibitem{DerbenevKondratenko1973}
Ya.~S. Derbenev and A.~M. Kondratenko, 
``Polarization kinetics of particles in storage rings'', 
Zh. Eksp. Teor. Fiz. \textbf{64}, 1918 (1973) 
[Sov. Phys.--JETP \textbf{37}, 968 (1973)].

\bibitem{Baier1971}
V.~N. Baier, 
``Thermal Excitations of Accelerated Electrons'', 
Usp. Fiz. Nauk \textbf{105}, 441 (1971) 
[Sov. Phys.--Usp. \textbf{14}, 695 (1972)].



\bibitem{Chen:1998kp}
P.~Chen and T.~Tajima,
``Testing Unruh radiation with ultraintense lasers,''
Phys. Rev. Lett. \textbf{83}, 256-259 (1999).

\bibitem{Martin-Martinez:2010gnz}
E.~Martin-Martinez, I.~Fuentes and R.~B.~Mann,
``Using Berry's phase to detect the Unruh effect at lower accelerations,''
Phys. Rev. Lett. \textbf{107} (2011), 131301
doi:10.1103/PhysRevLett.107.131301.

\bibitem{Stargen:2021vtg}
D.~J.~Stargen and K.~Lochan,
``Cavity Optimization for Unruh Effect at Small Accelerations,''
Phys. Rev. Lett. \textbf{129} (2022) no.11, 111303
doi:10.1103/PhysRevLett.129.111303.

\bibitem{Lochan:2019osm}
K.~Lochan, H.~Ulbricht, A.~Vinante and S.~K.~Goyal,
``Detecting Acceleration-Enhanced Vacuum Fluctuations with Atoms Inside a Cavity,''
Phys. Rev. Lett. \textbf{125} (2020), 241301
doi:10.1103/PhysRevLett.125.241301.

\bibitem{Deswal:2025cjw}
A.~Deswal, N.~Arya, K.~Lochan and S.~K.~Goyal,
``Time-Resolved and Superradiantly Amplified Unruh Effect,''
Phys. Rev. Lett. \textbf{135} (2025) no.18, 183601
doi:10.1103/6z1l-kkmk.


\bibitem{Unruh1981}
W.~G. Unruh, ``Experimental Black-Hole Evaporation?'', 
Phys. Rev. Lett. \textbf{46}, 1351--1353 (1981). 

\bibitem{Garay:1999sk}
L.~J.~Garay, J.~R.~Anglin, J.~I.~Cirac and P.~Zoller,
``Black holes in Bose-Einstein condensates,''
Phys. Rev. Lett. \textbf{85}, 4643-4647 (2000).


\bibitem{Hu:2018psq}
J.~Hu, L.~Feng, Z.~Zhang and C.~Chin,
``Quantum simulation of Unruh radiation,''
Nature Phys. \textbf{15} (2019) no.8, 785-789
doi:10.1038/s41567-019-0537-1.


\bibitem{Gooding:2020scc}
C.~Gooding, S.~Biermann, S.~Erne, J.~Louko, W.~G.~Unruh, J.~Schmiedmayer and S.~Weinfurtner,
``Interferometric Unruh detectors for Bose-Einstein condensates,''
Phys. Rev. Lett. \textbf{125} (2020) no.21, 213603
doi:10.1103/PhysRevLett.125.213603.

\bibitem{Tian:2022gfa}
Z.~Tian, L.~Wu, L.~Zhang, J.~Jing and J.~Du,
``Probing Lorentz-invariance-violation-induced nonthermal Unruh effect in quasi-two-dimensional dipolar condensates,''
Phys. Rev. D \textbf{106} (2022) no.6, L061701
doi:10.1103/PhysRevD.106.L061701.

\bibitem{Philbin:2007ji}
T.~G.~Philbin, C.~Kuklewicz, S.~Robertson, S.~Hill, F.~Konig and U.~Leonhardt,
``Fiber-optical analogue of the event horizon,''
Science \textbf{319}, 1367-1370 (2008).

\bibitem{Nation:2009xb}
P.~D.~Nation, M.~P.~Blencowe, A.~J.~Rimberg and E.~Buks,
``Analogue Hawking Radiation in a dc-SQUID Array Coplanar Waveguide,''
Phys. Rev. Lett. \textbf{103}, 087004 (2009).



\bibitem{Friis:2012cx}
N.~Friis, A.~R.~Lee, K.~Truong, C.~Sabin, E.~Solano, G.~Johansson and I.~Fuentes,
``Relativistic Quantum Teleportation with superconducting circuits,''
Phys. Rev. Lett. \textbf{110} (2013) no.11, 113602
doi:10.1103/PhysRevLett.110.113602.



\bibitem{Felicetti:2015kta}
S.~Felicetti, C.~Sab{\'\i}n, I.~Fuentes, L.~Lamata, G.~Romero and E.~Solano,
``Relativistic Motion with Superconducting Qubits,''
Phys. Rev. B \textbf{92} (2015) no.6, 064501
doi:10.1103/PhysRevB.92.064501
[arXiv:1503.06653 [quant-ph]].

\bibitem{Katayama2020} 
H.~Katayama, N.~Hatakenaka, and T.~Fujii, 
 ``Analogue Hawking radiation from black hole solitons in quantum Josephson transmission lines,''
\textit{Phys. Rev. D} \textbf{102}, 086018 (2020).


\bibitem{Katayama2025} 
H.~Katayama and N.~Hatakenaka, 
``Circular-Motion Fulling-Davies-Unruh Effect in Coupled Annular Josephson Junctions,''
\textit{Phys. Rev. Lett.} \textbf{135}, 046001 (2025).

\bibitem{Quach:2021vzo}
J.~Q.~Quach, T.~C.~Ralph and W.~J.~Munro,
``Berry Phase from the Entanglement of Future and Past Light Cones: Detecting the Timelike Unruh Effect,''
Phys. Rev. Lett. \textbf{129} (2022) no.16, 160401
doi:10.1103/PhysRevLett.129.160401.

\bibitem{Cheng2026} 
X.~Cheng, Y.~Li, Z.~Tian, X.~Zhao, X.~Qin, \& Y.~Lin, 
``Quantum simulation of oscillatory Unruh effect with superposed trajectories,''
Sci. China-Phys. Mech. Astron. \textbf{69}, 230411 (2026).

\bibitem{Luo:2025iur}
Z.~Luo, Y.~Li, X.~Zhao, Z.~Xie, Z.~Tian and Y.~Lin,
``Experimental Demonstration of the Timelike Unruh Effect with a Trapped-Ion System,''
[arXiv:2510.24163 [quant-ph]].






\bibitem{Fedichev:2003id}
P.~O.~Fedichev and U.~R.~Fischer,
``Gibbons-Hawking effect in the sonic de Sitter space-time of an expanding Bose-Einstein-condensed gas,''
Phys. Rev. Lett. \textbf{91}, 240407 (2003).




\bibitem{Fischer:2004bf}
U.~R.~Fischer and R.~Schutzhold,
``Quantum simulation of cosmic inflation in two-component Bose-Einstein condensates,''
Phys. Rev. A \textbf{70}, 063615 (2004)
doi:10.1103/PhysRevA.70.063615.






\bibitem{MIT-bag-original}
A.~Chodos, R.~L.~Jaffe, K.~Johnson, C.~B.~Thorn and V.~F.~Weisskopf,
``A New Extended Model of Hadrons,''
Phys. Rev. D \textbf{9}, 3471-3495 (1974).

\bibitem{Friis:2011yd}
N.~Friis, A.~R.~Lee, D.~E.~Bruschi and J.~Louko,
``Kinematic entanglement degradation of fermionic cavity modes,''
Phys. Rev. D \textbf{85} (2012), 025012.
\bibitem{Friis:2013eva}
N.~Friis, A.~R.~Lee and J.~Louko,
``Scalar, spinor, and photon fields under relativistic cavity motion,''
Phys. Rev. D \textbf{88} (2013) no.6, 064028.

\bibitem{DLMF}
NIST Digital Library of Mathematical Functions, \url{https://dlmf.nist.gov/}, Release 1.1.10 of 2023-09-15.

\bibitem{RefX}
V.~Toussaint, ``Exponential Suppression of the Unruh Effect and Geometric Enhancement in  a Fermionic Cavity QED Setup,'' [arXiv:2510.11460 [gr-qc]] (2025).





\bibitem{Aspelmeyer:2013lha}
M.~Aspelmeyer, T.~J.~Kippenberg and F.~Marquardt,
``Cavity Optomechanics,''
Rev. Mod. Phys. \textbf{86}, 1391 (2014).

\bibitem{Fedichev:2003bv}
P.~O.~Fedichev and U.~R.~Fischer,
``'Cosmological' quasiparticle production in harmonically trapped superfluid gases,''
Phys. Rev. A \textbf{69} (2004), 033602
doi:10.1103/PhysRevA.69.033602.

\bibitem{Tian:2017wij}
Z.~Tian, J.~Jing and A.~Dragan,
``Analog cosmological particle generation in a superconducting circuit,''
Phys. Rev. D \textbf{95} (2017) no.12, 125003
doi:10.1103/PhysRevD.95.125003.



\bibitem{Orlando:1999xxc}
T.~P.~Orlando, J.~E.~Mooij, L.~Tian, C.~H.~van der Wal, L.~S.~Levitov, S.~Lloyd and J.~J.~Mazo,
``Superconducting persistent-current qubit,''
Phys. Rev. B \textbf{60} 15398--15413  (1999).

\bibitem{Moodij:1999}
J.~E. Mooij, T.~P. Orlando, L. Levitov, L. Tian, C.~H. van~der~Wal, and S. Lloyd, ``Josephson persistent-current qubit,'' Science \textbf{285}, 1036--1039 (1999).
\bibitem{Schumacher:1995nrx}
B.~Schumacher,
``Quantum coding,''
Phys. Rev. A \textbf{51} (1995) no.4, 2738.

\bibitem{NakamuraPashkinTsai1999}
Y.~Nakamura, Y.~Pashkin, and J.~S. Tsai, ``Coherent control of macroscopic quantum states in a single-Cooper-pair box,''
Nature \textbf{398}, 786--788 (1999).

\bibitem{Krantz:2019jkw}
P.~Krantz, M.~Kjaergaard, F.~Yan, T.~P.~Orlando, S.~Gustavsson and W.~D.~Oliver,
``A quantum engineer's guide to superconducting qubits,''
Appl. Phys. Rev. \textbf{6} (2019) no.2, 021318.

\bibitem{DevoretWallraffMartinis2004}
M.~H. Devoret, A. Wallraff, and J.~M. Martinis, ``Superconducting qubits: A short review,''
\textit{arXiv preprint arXiv:cond-mat/0411174 [cond-mat.mes-hall]} (2004).

\bibitem{Paauw2009}
F.~G. Paauw, {\em Superconducting flux qubits: Quantum chains and tunable qubits},\\
Doctoral Thesis, Delft University of Technology, Delft, Netherlands, 2009.\\
Promotors: J.~E. Mooij and C.~J.~P.~M. Harmans.\\
\url{https://resolver.tudelft.nl/uuid:9ed11bb5-0d00-4d33-96f7-b7a6c60d79b2}




\bibitem{Wendin:2017ymy}
G.~Wendin,
``Quantum information processing with superconducting circuits: a review,''
Rept. Prog. Phys. \textbf{80} (2017) no.10, 106001.

\bibitem{Blais2005}
A.~Blais, R.-S. Huang, A.~Wallraff, S.~M. Girvin, and R.~J. Schoelkopf, ``Circuit quantum electrodynamics,''
Rev. Mod. Phys. \textbf{77}, 1225--1260 (2005).

\bibitem{Koch2007}
J.~Koch, T.~M. Yu, J.~Gambetta, A.~A. Houck, D.~I. Schuster, J.~Majer, A.~Blais, M.~H. Devoret, S.~M. Girvin, and R.~J. Schoelkopf,
``Charge-insensitive qubit design derived from the Cooper pair box,''
\textit{Phys. Rev. A} \textbf{76}, 042319 (2007).

\bibitem{CastellanosBeltran2008}
M.~A. Castellanos-Beltran, K.~D. Irwin, G.~C. Hilton, L.~R. Vale, and J.~A. B. Jr.,
``Low-loss superconducting transmission line resonators with high internal quality factors,''
\textit{Appl. Phys. Lett.} \textbf{93}, 123508 (2008).

\bibitem{Pozar2011}
D.~M. Pozar, \textit{Microwave Engineering}, 4th ed. (Wiley, Hoboken, NJ, 2011).




\bibitem{ClarkeWilhelm2008}
J.~Clarke and F.~K. Wilhelm, ``Superconducting quantum bits,''
\textit{Nature} \textbf{453}(7198), 1031--1042 (2008).

\bibitem{Rabi:1936lvg}
I.~I.~Rabi,
``On the Process of Space Quantization,''
Phys. Rev. \textbf{49} (1936) no.4, 324.

\bibitem{JaynesCummings1993}
E.~T. Jaynes and F.~W. Cummings, ``Comparison of quantum and semiclassical radiation theories with application to the beam maser,''
Proc. IEEE \textbf{51}, 89--109 (2004).

\bibitem{Braak:2011tmj}
D.~Braak,
``Integrability of the Rabi Model,''
Phys. Rev. Lett. \textbf{107} (2011) no.10, 100401.


\bibitem{ShoreKnight1993}
B.~W. Shore and P.~L. Knight, ``The Jaynes-Cummings model,''
 Journal of Modern Optics \textbf{40}, 1195--1238 (1993).

\bibitem{GerryKnight2005}
C.~C. Gerry and P.~L. Knight, \textit{Introductory Quantum Optics} (Cambridge University Press, Cambridge, England, 2005).



\bibitem{Fedichev:2003dj}
P.~O.~Fedichev and U.~R.~Fischer,
``Observer dependence for the phonon content of the sound field living on the effective curved space-time background of a Bose-Einstein condensate,''
Phys. Rev. D \textbf{69} (2004), 064021
doi:10.1103/PhysRevD.69.064021.

\bibitem{AbramowitzStegun1965}
M. Abramowitz and I. A. Stegun (Eds.), Handbook of Mathematical Functions with Formulas, Graphs, and Mathematical Tables, Dover Publications, New York, 1965.


\end{thebibliography}
\end{document}